# Electronic Structure of the Dark Surface of the Weak Topological Insulator $Bi_{14}Rh_3I_9$

*Christian Pauly[1], Bertold Rasche[2], Klaus Koepernik[4,5], Manuel Richter[4,5], Sergey Borisenko[4], Marcus Liebmann[1], Michael Ruck[2,3], Jeroen van den Brink[4,5], and Markus Morgenstern[1]*

[1]II. Institute of Physics B and JARA-FIT, RWTH Aachen University, D-52074 Aachen, Germany. [2]Department of Chemistry and Food Chemistry, TU Dresden, D-01062 Dresden, Germany. [3]Max Planck Institute for Chemical Physics of Solids, D-01187 Dresden, Germany. [4]Leibniz Institute for Solid State and Materials Research, IFW Dresden e.V., PO box 270116, D-01171 Dresden, Germany. [5]Dresden Center for Computational Materials Science (DCMS), TU Dresden, D-01069 Dresden, Germany

ABSTRACT: The compound $Bi_{14}Rh_3I_9$ consists of ionic stacks of intermetallic $[(Bi_4Rh)_3I]^{2+}$ and insulating $[Bi_2I_8]^{2-}$ layers and has been identified to be a weak topological insulator. Scanning tunneling microscopy revealed the robust edge states at all step edges of the cationic layer as a topological fingerprint. However, these edge states are found 0.25 eV below the Fermi level which is an obstacle for transport experiments. Here, we address this obstacle by comparing results of density functional slab calculations with scanning tunneling spectroscopy and angle-resolved photoemission spectroscopy. We show that the n-type doping of the intermetallic layer is intrinsically caused by the polar surface and is well screened towards the bulk. In contrast, the anionic "spacer" layer shows a gap at the Fermi level, both, on the surface and in the bulk, *i.e.* it is not surface-doped due to iodine desorption. The well screened surface dipole implies that a buried edge state, probably already below a single spacer layer, is located at the Fermi level. Consequently, a multilayer step covered by a spacer layer could provide access to the transport properties of the topological edge states. In addition, we find a lateral electronic modulation of the topologically non-trivial surface layer which is traced back to the coupling with the underlying zigzag chain structure of the spacer layer.

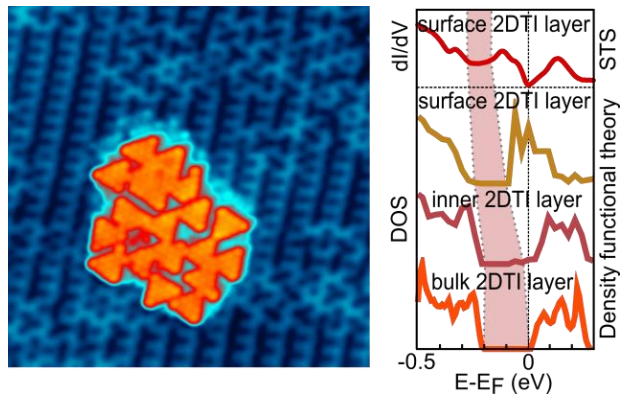

Table of content figure: left: scanning tunneling microscopy image of cleaved $Bi_{14}Rh_3I_9$ with topological layer (blue) and trivial spacer layer (orange); right: local density of states resulting from experimental dI/dV curve on the topological layer called 2DTI (top) and from density functional theory calculations of the layers as marked (bottom)

states with a helical spin-momentum locking. The latter suppresses backscattering which makes these materials interesting for quantum devices.[1-4] Three-dimensional (3D) TIs are characterized by a set of four topological indices, namely $v_0$; ($v_1$, $v_2$, $v_3$) with $v_i = \{0, 1\}$.[5] If $v_0 = 1$, the system is called a strong 3DTI and has robust metallic surface states on all surfaces. If $v_0 = 0$ and any other $v_i \neq 0$, the system is called a weak 3DTI with robust metallic surface states on some, but not all surfaces.[2,6,7] Whereas the current research is focused on strong 3DTIs, *e.g.* $Bi_2Se_3$,[8-10] weak 3DTIs have barely been probed. This is most likely due to the initial believe that their topological nature with two Dirac cones at one surface would be unstable with respect to any kind of disorder.[2] However, recent theoretical work has clarified that the topological surface states of a weak 3DTI are instead stabilized by disorder which counteracts the destabilizing dimerization of adjacent two-dimensional (2D) TI layers.[11-15]

The compound $Bi_{14}Rh_3I_9$ was recently synthesized and characterized by X-ray diffraction (XRD) to be a layered ionic structure with alternating *cationic* $[(Bi_4Rh)_3I]^{2+}$ and *anionic* $[Bi_2I_8]^{2-}$ layers (Figure 1a and 1b).[16,17] The former has been identified as a 2DTI sheet and the latter as an insulating spacer layer using density functional theory (DFT) calculations. The stacked layer system is a weak 3DTI according to DFT yielding $v_0$; $(v_1, v_2, v_3) = 0$; $(0, 0, 1)$.[16] The electronic band-structure from DFT is thereby in agreement with angle-resolved photoemission spectroscopy (ARPES).[16] More recently, scanning tunneling spectroscopy (STS) on the topologically dark (001) surface of single crystalline $Bi_{14}Rh_3I_9$ confirmed the predicted weak TI properties by revealing a one-dimensional (1D) state localized at all step edges of the *cationic* 2DTI surface layer throughout the whole topologically non-trivial band gap. These edge states barely show backscattering, in remarkable contrast to conventional 1D edge states.[18] Thus, STS confirms one of the hallmarks of the weak 3DTI properties.[19] Moreover, the edge states are only 1 nm wide, which could allow the construction of devices with extremely small active elements. Scratching with a sharp needle has been proposed as an easy way to construct the conducting edge channels artificially.[18]

However, both ARPES and STS reveal that the non-trivial band gap of a width of ~ 150 meV has its center at approximately 250 meV below the Fermi level $E_F$.[16,18] Thus, both techniques observe a considerable n-type doping of the $Bi_{14}Rh_3I_9$ (001) surface in comparison to the bulk DFT band structure calculations, which find the non-trivial gap at $E_F$.[16] A possible reason for this intrinsic doping is arising from the fact that the surface is polar. Consequently, the surface layer is less charged than a bulk layer with the same composition, a behavior which is well known, *e.g.*, for polar oxide surfaces.[20] In order to support the hypothesis of intrinsic surface doping, we compare experimental electronic structure data of the two distinct layer types as recorded by STS with a finite-slab DFT calculation using a structure model consisting of four layers (two of each type). The DFT calculations reproduce the energy shift of the topologically non-trivial band gap at the surface as found in the experiment. Based on these results, we suggest a strategy to compensate the surface charge for $Bi_{14}Rh_3I_9$ and similar compounds[20] and, thus, to move the gap with the topological edge state towards $E_F$.

In addition, we find that the atomic appearance of the graphene-like surface layer in scanning tunneling microscopy (STM) images is partly different from the perfect honeycomb lattice expected for the 2DTI layer. We observe a rectangular $\sqrt{3}\,a_0 \times a_0$ type unit cell instead of a rhombic unit cell with unit vector $a_0$, *i.e.* the area of the unit cell is doubled. The appearance of the 2DTI layer moreover is strongly dependent on bias polarity showing that it is given by an electronic effect. By comparison with real space density of states images deduced from the DFT slab calculation, we suggest an interlayer coupling with the zigzag chains of the underlying anionic layer being responsible for the modulation of the electronic structure of the surface 2DTI layer.

RESULTS AND DISCUSSION:

A large-scale STM topography image of the freshly cleaved $Bi_{14}Rh_3I_9$ (001) is shown in Figure 1c revealing two different layers which are identified as the cationic 2DTI layer and the anionic spacer layer by their different step heights (inset).[18] A more straightforward distinction is their different atomic appearance as shown in the inset of Figure 2a and 2b. Figure 1d shows the surface electronic structure of the cleaved $Bi_{14}Rh_3I_9$ as the momentum distribution of photoemission intensity integrated within the 0.5 eV energy interval below $E_F$. At the chosen photon energy ($hv$ = 80 eV), the photoemission experiment is sensitive to the top layer only (few Å).[22] A distorted honeycomb pattern is visible as expected for a crystal with a triclinic symmetry stemming from a highly symmetric 2DTI honeycomb lattice and a lower-symmetric spacer layer.[16, 17] In order to reveal the band gap position of the $Bi_{14}Rh_3I_9$ surface, we plot the energy distribution curve (EDC) in Figure 1e as determined by integrating over the whole $k_x < 0$ range from Figure 1d. A gap-like feature is discernible between -170 meV and -370 meV (marked by the shaded area). This band gap area has been identified previously by E(**k**)-plots and by STS.[16,18] We checked the accuracy of the EDC and its gap position by integrating over different portions of the $k$-space, always resulting in a similar distribution.

Whereas ARPES measures the band structure properties integrated over a larger length scale (100 μm), thus mixing the electronic features of the 2DTI and the spacer layer, STS is able to measure the local density of states (LDOS) on the atomic scale by recording the differential conductivity $dI/dV$ with respect to the sample voltage $V$.

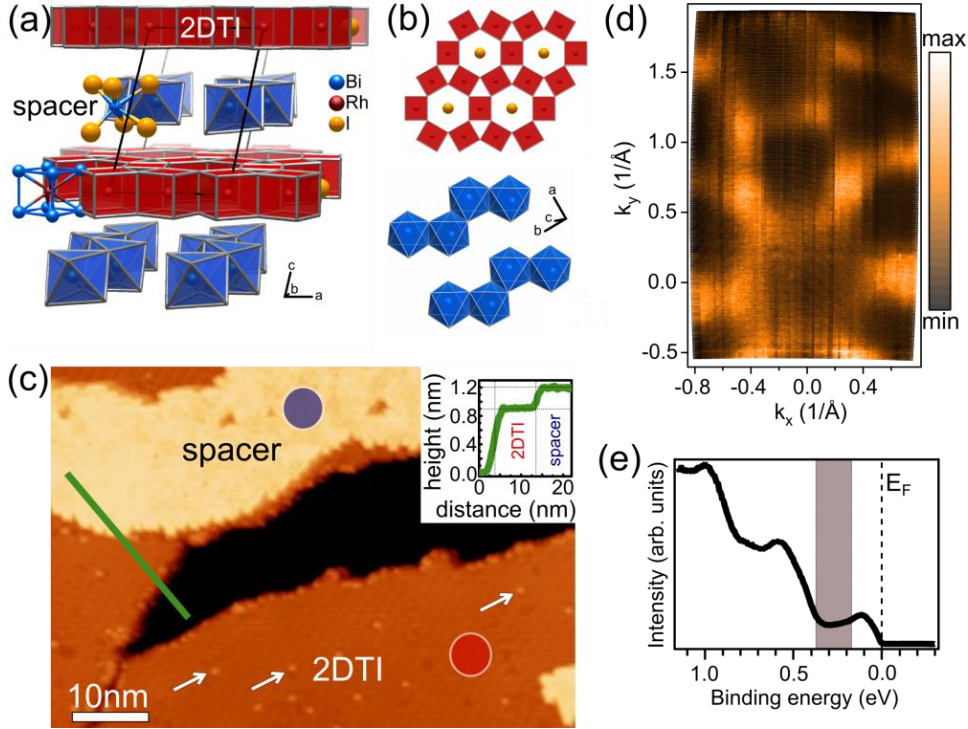

**Figure 1.** (a) Sketch of $Bi_{14}Rh_3I_9$ as deduced from XRD. The sketch consists of two cationic 2DTI layers $[(Bi_4Rh)_3I]^{2+}$ (red) and two anionic spacer layers $[Bi_2I_8]^{2-}$ (blue). This particular geometry is used for the finite slab calculation of Figure 2 and 4. (b) Top view of the 2DTI (top, red) and spacer layer (bottom, blue). (c) STM image of cleaved $Bi_{14}Rh_3I_9$ ($V$ = 0.8 V, $I$ = 80 pA) revealing three different layers. The upper two layers are identified as the 2DTI and spacer layer by their different step heights (see inset) according to Pauly *et al.*[18] They are marked accordingly. The red and blue dot indicate the measurement positions of the STS spectra from Fig. 2(a) and (b), respectively, and the arrows mark adatoms on the 2DTI layer. (d) Momentum distribution of ARPES intensity integrated between 0.5 eV binding energy and $E_F$, $h\nu$ = 80 eV. (e) Energy distribution curve from ARPES integrated over several Brillouin zones (the whole ($k_x < 0$) area of panel (d)). Shaded area marks the non-trivial gap.

The corresponding curves for the 2DTI and spacer layer are shown in Figure 2a and 2b, respectively. The bottom spectra in both figures (dark red and dark blue) are recorded at the positions marked by the respective color in Figure 1c. The other two spectra are measured elsewhere with the same microtip. The $dI/dV(V)$ spectra measured on the 2DTI layer reveal the non-trivial band gap [18] (light-red shaded area in Figure 2a) between $V$ = -170 mV and $V$ = -290 mV for the bottom curve, respectively between $V = $ -240 mV and $V = $ -370 mV for the top curve. Both cases exhibit a considerable n-type doping of the 2DTI surface layer in line with the EDC from the ARPES. The slight shift in energy between the two curves is most probably due to local potential fluctuations induced by charge transfer from surface adatoms or from defects within adjacent spacer layer areas, see inset in Figure 2 b. Adatoms are indeed visible as bright dots marked by arrows on the 2DTI layer in Figure 1c. Most likely, the adatoms are remaining iodine atoms from the cleavage process, since their density does not change by increasing the waiting time between cleavage and cool-down by a factor of ten. The low adatom density, however, cannot account for the strong doping. Moreover, iodine atoms would attract electrons leading to p-type doping instead of the observed n-type doping. The potential fluctuations anyway explain the larger effective band gap observed by ARPES with respect to the STS band gap, since ARPES averages over a large surface area.

Note, that the electronic structure of the 2DTI layer exhibits vanishing $dI/dV$ intensity at $V = 0$ V ($E_F$) surrounded by a nearly linear increase of the LDOS. This type of LDOS suppression is attributed to a Coulomb pseudo-gap of Efros-Shklovskii type originating from electron-electron interaction which reduces the density of states around $E_F$ almost linearly for two-dimensional (2D) electron systems.[23,24] This feature, thus, points to a quasi-2D nature of the surface 2DTI layer and hence to a weak interaction with the underneath spacer layer, at least, close to $E_F$.

The $dI/dV(V)$ spectra of the anionic spacer layer reveal a larger band gap (light-blue shaded area in Figure 2b) located between $V \approx 0$ mV and $V \approx $ -310 mV. As in the case of the 2DTI layer, small energy shifts between the two curves are discernible. Notice, that the gap of the spacer layer covers the major part of the gap of the 2DTI surface layer.

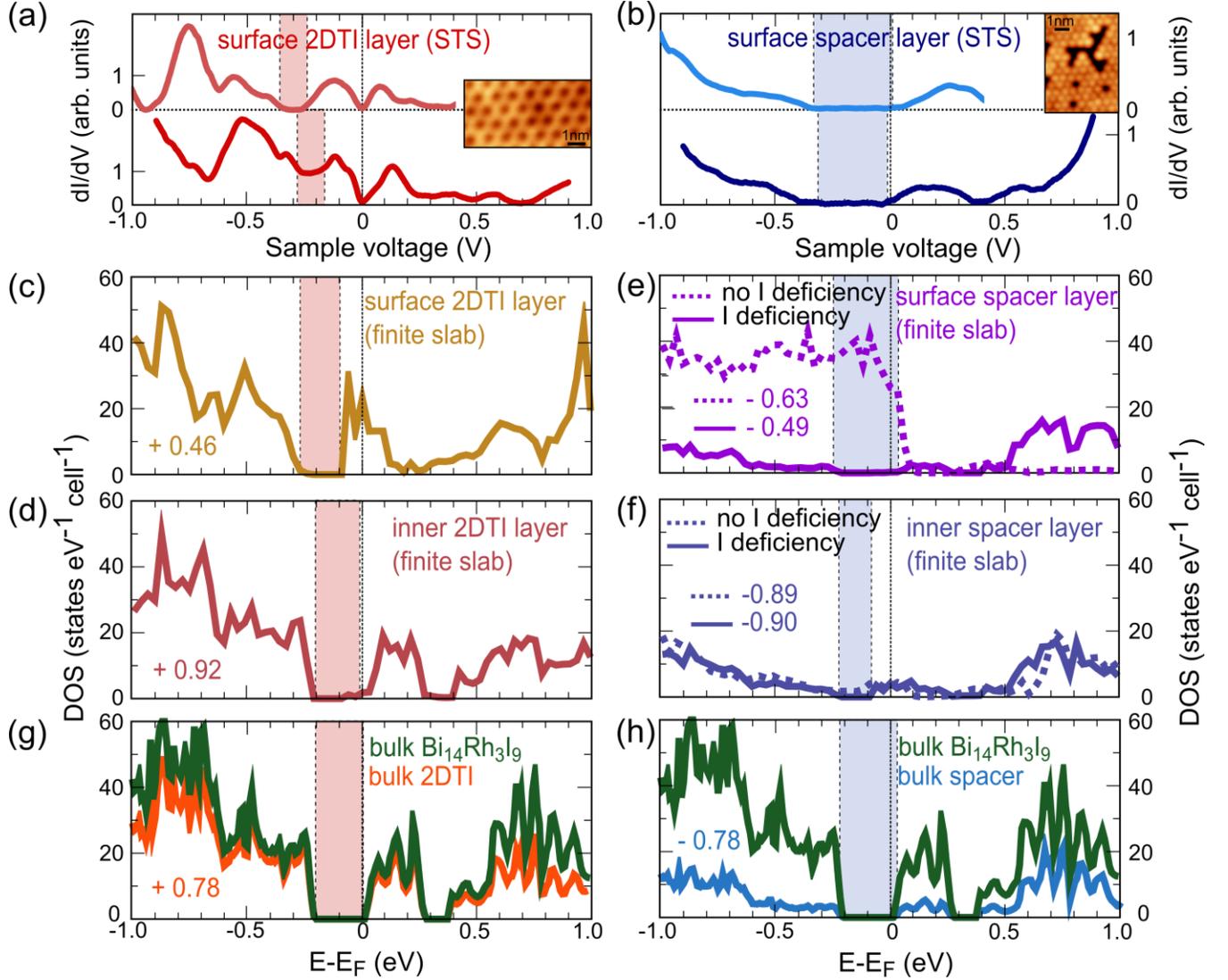

**Figure 2.** (a) Bottom: $dI/dV(V)$ spectrum of the 2DTI layer taken at the position marked in Figure 1c by a red dot ($V_{stab} = 1$ V, $I_{stab} = 70$ pA, $V_{mod} = 4$ mV). Top: Additional $dI/dV(V)$ spectrum recorded on a different 2DTI layer on the same sample surface ($V_{stab} = 1$ V, $I_{stab} = 200$ pA, $V_{mod} = 4$ mV). Light red area marks the band gap region. Inset: Atomically resolved STM image of the 2DTI layer exhibiting a completely intact honeycomb lattice ($V = +1.5$ V, $I = 100$ pA). (b) Same as (a) for the spacer layer with the $dI/dV(V)$ spectrum on the bottom ($V_{stab} = 1$ V, $I_{stab} = 70$ pA, $V_{mod} = 4$ mV) recorded at the position marked blue in Figure 1c and additional spectrum (top) recorded elsewhere ($V_{stab} = 1$ V, $I_{stab} = 200$ pA, $V_{mod} = 4$ mV). Light blue area marks the band gap region. Inset: Atomically resolved STM image of the spacer layer revealing an iodine deficiency, which is visible as missing atoms in the hexagonal structure amounting to ~20% of the iodine atoms in the outermost layer ($V = -1.3$ V, $I = 100$ pA). (c)-(f) Layer-resolved DOS of surface as well as inner 2DTI and spacer layer as marked, deduced from the four-layer slab calculation using the structure in Figure 1a. For the full lines in (e), (f) one iodine atom per formula unit is effectively

removed from the outer part of the surface Bi-I layer in agreement with the experimental observation of missing iodine atoms. The removal is modeled by a VCA-type calculation. The dotted lines in (e) and (f) mark the DOS of the surface and inner spacer layer as calculated using an intact spacer without iodine removal. (g), (h) Calculated total DOS of bulk $Bi_{14}Rh_3I_9$ (green curve) and of a single 2DTI layer (orange curve in (g)), respectively a single spacer layer (light blue curve in (h)) using the bulk structure. The numbers in (c)-(h) are Mulliken-like gross charges per formula unit of the related layers and the shaded areas mark the approximate gap region of the respective spectra.

Figure 2c and 2e show the surface electronic structure deduced from the finite slab calculation of the (001) surfaces of $Bi_{14}Rh_3I_9$. By using a stack with different top- and bottom-layers (Figure 1a), both types of surface layers are studied in one and the same calculation. The top surface of the slab is the 2DTI layer, while the bottom surface is the spacer layer, which are both bordered by vacuum. Additionally, there is an inner 2DTI layer sandwiched between two adjacent spacer layers and an inner spacer layer sandwiched between two 2DTI layers. The latter two arrangements model the situation as found in the infinite bulk structure. Using this structure, two slab calculations were performed. In the first calculation, only electronic relaxation at the two surfaces was considered. In the second calculation, a virtual crystal approximation (VCA) was applied in order to account for the experimentally observed iodine deficiency. We assume an iodine deficiency of ~ 25 % for the outermost iodine atomic layer of the surface spacer layer (see Figure 1a) in accordance with experiment and chemical reasoning (see below). The outermost iodine layer of the spacer is therefore replaced by a layer of iodine-like atoms with the non-integer atomic charge $Z = 53.25$. In this way, the effect of desorption of every fourth atom of the outermost iodine layer on the electron number is simulated without introducing disorder. The bottom iodine atomic layer, which is adjacent to the 2DTI layer (Figure 1a), however, is kept intact, such that one of eight iodine atoms per formula unit is effectively removed. The surface 2DTI layer, in contrast, is modelled in its bulk structure, since no signs of desorption have been observed in STM images (see inset in Figure 2a).

Before discussing the obtained electronic structures, we provide arguments that two double-layers are a sufficient model to describe the polar surfaces of $Bi_{14}Rh_3I_9$. It is known, that the macroscopic component of the surface dipole moment at a polar surface has to be compensated by electronic and/or structural reconstruction with related charge modification at the surface.[20] The fact that no structural reconstruction is observed at the 2DTI surface can be explained by relatively strong bonds within this layer and a high polarizability of the layer leading to an effective screening without structural reconstruction. The high polarizibility is deduced from the large density of states at the lower edge of the conduction band (Figure 2g). On the contrary, the spacer layer has a low polarizability (low density of states at the upper edge of the valence band, Figure 2h) and presumably weaker bonds. Thus, removal of atoms (iodine) is energetically more favorable for the spacer layer in order to compensate the surface dipole.

Electrostatic considerations show that, for the present case of equidistant layers with alternating bulk charges $\pm Q$, the macroscopic dipole moment vanishes in the limit of an infinitely thick slab, if the charge $Q_i$ on the outer layers $i$ fulfills [20]:

$$\sum_i^m Q_i = \pm (-1)^{m+1} Q/2,$$

where the − (+) sign stands for the case of an outermost anionic (cationic) layer. A decay of the charge modification within a finite number of layers $m$ is thereby assumed.

Table 1 summarizes our results of layer-wise Mulliken analysis for the bulk and slab charges. We note that the bulk layer charges ($|Q| = 0.78$) amount to only ~40% of the nominal charges ($|Q_{nom}|$ = 2). The value of $Q$ depends, of course, on the specific projection onto atomic–like basis states, see *e.g.* Refs.[20,21]. Comparing the dipole moments obtained for a slab cut from the bulk without electronic relaxation, a slab with electronic relaxation, and slab with iodine deficiency, we find a ratio of about 1: 0.5: 0.3 by assuming the charge per layer to be concentrated in the central plane of the respective layer. This shows that the electronic and structural relaxation considered in the four-layer model are effective in reducing the surface-induced dipole moment, albeit not screening it completely. In the model with iodine deficiency, the major contribution to the compensation takes place in the outermost layers only.

An important result of this analysis is the total size of the compensating charge, obtained by summation over two respective layers: $\sum_i^2 Q_i = \pm 0.39$ ($\pm 0.44$) for the slab without (with) iodine deficiency. These numbers deviate by less than 13% from the ideal value of $\pm 0.39$ that is expected for a slab of infinite thickness. This means that the band filling and related electronic structure details of the considered slab models should almost match those of slabs in the limit of infinite thickness.

TABLE 1. Mulliken gross charges $Q$ and $Q_i$ for bulk, slab and slab with iodine deficiency (see text). The charges are given in units of the elementary charge $e$ per formula unit.

| Layer | $Q$ (bulk) | $Q_i$ (slab) | $Q_i$ (slab with iodine deficiency) |
|---|---|---|---|
| surface 2DTI | +0.78 | +0.50 | +0.46 |
| inner spacer | -0.78 | -0.89 | -0.90 |
| inner 2DTI | +0.78 | +1.02 | +0.92 |
| surface spacer | -0.78 | -0.63 | -0.49 |

The calculated density of states (DOS) together with the Mulliken charges of the related layers are shown for the 2DTI layers in Figure 2c and 2d, and for the spacer layers in Figure 2e and 2f.

For comparison, the DOS of the spacer layers from the slab model without iodine deficiency is plotted as dotted lines in Figure 2e and 2f.

The surface 2DTI layer (Figure 2c) exhibits a half-filled first conduction band with the band gap lying between -280 meV and -100 meV below $E_F$. This is in reasonable agreement with the gap position found in the STS and the ARPES spectra (Figure 2a and 1e). In contrast, the inner 2DTI layer is found to have a charge of +0.92 which is close to the charge of a bulk 2DTI layer (+0.78). Moreover, the band gap of the inner 2DTI layer is shifted upwards by about 100 meV with respect to the surface layer, such that $E_F$ is approximately at the band gap edge. For comparison, Figure 2g shows the DOS of a bulk 2DTI layer (orange curve). The good matching between Figure 2d and 2g suggests that the surface dipole is well screened by the additional charge in the surface 2DTI conduction band.

The DOS of the surface spacer layer including the removal of one iodine (continuous line in Figure 2e), reveals a larger band gap than the 2DTI layer. Its width is $\Delta E \approx 300$ meV with $E_F$ located at the upper edge of the band gap. Both, width and position of the gap are in accordance with the STS data (Figure 2b) and, in addition, are very similar to the spacer layer within the bulk calculation (Figure 2h). In contrast, the DOS of the surface spacer layer without iodine removal (dotted line in Figure 2e) is strongly shifted and distorted with respect to the bulk spacer layer DOS and does not match the experiment. Figure 2f shows the corresponding DOS for the inner spacer layer. Apart from a smaller band gap size, no obvious energy shift of the electronic structure is discernible with respect to the bulk layer (Figure 2h). Moreover, the inner spacer layer is hardly affected by the removal of iodine atoms from the surface spacer layer (Figure 2f). Thus, the surface dipole of the spacer layer is well screened by either pure electronic or combined structural and electronic relaxation. Only the latter, however, yields an electronic structure that agrees with the measured spectra.

We also evaluated the DOS of the surface and inner 2DTI layer for the calculation without iodine removal (not shown here). Almost no change is found for the surface 2DTI layer and only a small deviation for the inner 2DTI layer with respect to the spectra in Figure 2c and 2d.

The n-doping and the iodine removal as done for the modeling can be rationalized by retracing the mechanical cleavage process of the layered $Bi_{14}Rh_3I_9$ crystal. The STM investigations of surfaces generated by cleavage under UHV conditions show that the intermetallic layer $[Bi_{12}Rh_3I]^{2+}$ typically remains undamaged while the spacer $[Bi_2I_8]^{2-}$ is fragmented in various ways. Presuming that the two crystal fragments bear no overall charge, two borderline scenarios have to be considered.

In the case of homogeneous splitting, the spacer is bisected and two $[BiI_4]^{1-}$ surfaces are generated. Here, the Fermi level is expected to stay in the topologically non-trivial energy gap. Unfortunately, this variant is little probable due to the bonding strength within the spacer layer and does not meet our observations.

Alternatively, the crystal cleaves between the intermetallic layer and the spacer, which then form the new surfaces of the two fragments (heterogeneous splitting). According to our assumption of uncharged fragments, the terminating intermetallic layer will have the nominal charge of +1 instead of +2 per formula unit due to the missing anionic spacer layer on top (chemically reduced, n-doped surface). The additional electron is accommodated by shifting the conduction band edge below the Fermi level, in this way rendering the dark face of the TI metallic.

The terminating spacer of the other fragment necessarily has one electron less than needed. Such an oxidized (p-doped) $[Bi_2I_8]^{1-}$ layer is chemically unstable and decomposes into $[Bi_2I_7]^{1-}$ + ½ $I_2$, such that nominally favorable $Bi^{3+}$ and $I^-$ configurations are realized. As we believe that only the upper iodine layer of the surface spacer layer (Figure 1a) desorbs iodine, the iodine deficiency in that layer would nominally amount to 25 %. Desorption of molecular iodine is supported by the UHV conditions under which the cleavage was performed. After this chemical (redox) reaction, which removes the p-doping of the terminating layer, the Fermi level is situated in the band gap.

Indeed, the band gap is shifted downwards by the iodine removal (continuous line in Figure 2e) with respect to the gap position of an unreconstructed $[Bi_2I_8]^{1-}$ surface spacer layer (dotted line in Figure 2e). In addition, the unreconstructed layer shows a large intensity at and below $E_F$ in contrast to the experiment, which we attribute to the DOS of the formally neutral I atom. This large DOS is the electronic fingerprint of the discussed structural instability which facilitates the observed desorption process of iodine during the cleavage.

Experimentally, the iodine deficiency in the surface spacer layer has been determined by simply counting the missing atoms in STM images. Therefore, we overlay the atomic model structure of a perfect spacer lattice. The deficiency of iodine atoms is found to be on average 19 ± 3 %. However, one has to keep in mind that we can only count missing iodine atoms in the interior of the spacer layer, while iodine atoms desorbed from the edge cannot be distinguished from the edge itself. Thus, the measured deficiency is well in line with the scenario of charge compensation by 25 % iodine desorption. Alternatively, the edge of the spacer layer surface can be interpreted as a homogeneous splitting scenario, such that iodine desorption takes only place in the interior of the spacer.

As a conclusion of this discussion, the good agreement between the DFT and the experimental data in terms of band gap and iodine deficiency suggests that the spacer layer, in contrast to the 2DTI layer, is able to adapt its chemical composition, *i.e.* to compensate the surface dipole by removing iodine atoms.

As discussed, the overall surface should exhibit areas with $E_F$ in the gap (spacer after iodine removal) and $E_F$ within the conduction band (2DTI). This seems to be in contradiction to our previous finding,[16, 22] that the ARPES band structure matches very well the DFT band structure projected onto the (n-doped) 2DTI surface. Here, we note that for the irreversibly reconstructed surface spacer layer, the DOS below and above the gap (between -0.5 eV and 0.5 eV) is very low, such that the contribution from spacer surface areas to ARPES intensities is marginal. Hence, though both possible terminations are in the focus of the incoming photons, only the 2DTI termination essentially contributes to the ARPES intensity around $E_F$.

It is accepted that a weak 3DTI can be assembled as a stack of 2DTIs. We have shown before, that the free-standing, relaxed cationic $[(Bi_4Rh)_3I]^{2+}$ layer in its nominal charge state is indeed a 2DTI.[16] No information is available, hitherto, about the free-standing spacer layer. Thus, we additionally calculated the topological invariants of the spacer layer by using a single $[Bi_2I_8]^{2-}$ layer after relaxation within the layer symmetry *p2/c* and with cell constants according to the experiment. The VCA method has been applied in order to account for the charge: All iodine atoms are replaced by iodine-like atoms with Z = 53.25. It turns out that the spacer layer is a trivial 2D insulator with topological indices 0; (0, 0, 0) and a band gap of 1.71 eV. The gap is larger than, *e.g.* in the surface spacer layer deduced from the slab calculation (continuous line in Figure 2e), most probably due to the fact that no adjacent layers are included.

Next, we investigate the coupling of the surface 2DTI layer to the underlying, trivial spacer layer. In general, for stacked weak 3DTIs, translation-invariant coupling between the layers cannot gap out the two topological Dirac cones of the bright surfaces, as long as the bulk band order is not changed by the coupling.[11] However, the strength of the interlayer coupling affects the dispersion of the Dirac cones. For weak interlayer coupling, as for KHgSb, proposed to be a weak 3DTI on the base of DFT calculations, the band dispersion of the Dirac cones along the stacking direction is small.[26] In contrast, for strong interlayer coupling, as for Bi$_2$TeI, proposed to be a stacked weak 3DTI by DFT, the DFT calculations reveal largely isotropic Dirac cones.[27] For the latter material, the strong interlayer coupling is even critical for the topological indices, *i.e.* the coupling delivers the required inversion of bulk bands. In contrast, DFT calculations of Bi$_{14}$Rh$_3$I$_9$ [16] and KHgSb [26] demonstrate that the band inversion occurs within the 2DTI layer itself, whereas the topologically trivial layers (*e.g.* the Bi-I layer in Bi$_{14}$Rh$_3$I$_9$) only act as an intercalation separating adjacent 2DTIs.[18]

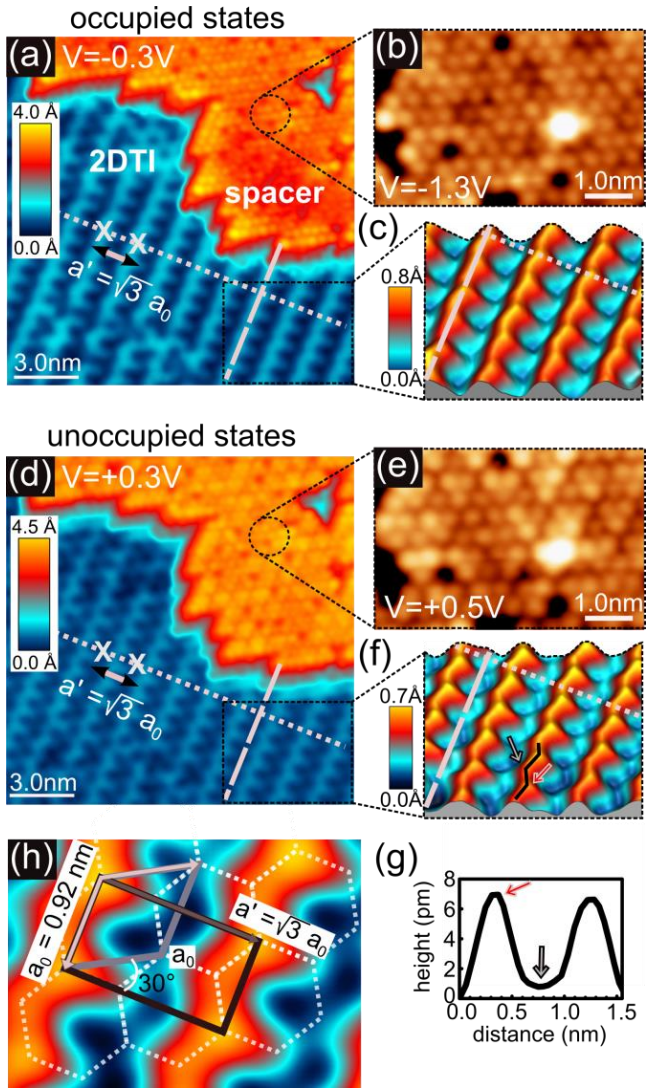

**Figure 3.** (a), (d) STM image of the 2DTI and spacer layer (as marked in (a)) recorded at different bias voltages, (a) occupied states ($V = -0.3$ V, $I = 80$ pA), (d) unoccupied states ($V = +0.3$ V, $I = 80$ pA). Gray dotted and dashed lines in both images mark the same sample positions. The distance between adjacent atomic rows on the 2DTI layer is marked by two crosses and is denoted as a'. The step edge shape of the spacer layer serves as a reference point for the gray lines. (b), (e) Atomically resolved STM image of the spacer layer measured for the occupied states (b) ($V = -1.3$ V, $I = 200$ pA) and unoccupied states (e) ($V = +0.5$ V, $I = 200$ pA), showing slightly different atomic appearance. (c), (f) Zoom and 3D representation of the 2DTI layer regions as marked by the black dotted rectangles in (a), (d), respectively. The black zig-zag line in (f) marks the direction of the profile line in (g) with arrows marking positions. (g) Height profile along the black zig-zag line marked in (f) with arrows at the same positions as in (f) marking extrema of the profile line. (h) 2D representation and further zoom into (f) highlighting the rectangular $\sqrt{3}\, a_0 \times a_0$ appearance of the 2DTI layer by overlaying the honeycomb lattice (dotted lines), the rhombic

unit cell of the honeycomb lattice (gray full lines with indicated unit vectors $a_0$), and the apparent unit cell of the STM image (black lines with unit vectors $a_0$ and $a' = \sqrt{3}\, a_0$).

Figure 3a and 3d show two STM images of an identical area probed at negative (occupied states) and positive (unoccupied states) sample bias, respectively. The area includes a spacer layer on top of the 2DTI layer as marked in Figure 3a. Interestingly, the STM image reveals an arcade-type pattern for the 2DTI layer which strongly deviates from the honeycomb symmetry of the 2DTI layer structure (Figure 1b). The unperturbed honeycomb structure with its rhombic unit cell ($a_0 = 0.92$ nm [16]) is, however, visible with other tips as in the inset of Figure 2a. The apparent structure in Figure 3a and 3d looks as if one would map only one half of the honeycombs, which becomes most obvious within the zoom images in Figure 3c, 3f, and 3h.

Moreover, by comparing the images of occupied and unoccupied states, a shift of the row pattern by $\frac{\sqrt{3}}{2} a_0$ perpendicular to the visible rows and $\frac{1}{2} a_0$ in parallel direction is discernible. This corresponds to mapping one and the other half of the honeycombs in occupied and unoccupied states, respectively (Figure 3h). The lateral shift gets even more obvious by comparing the zoomed images (Figure 3c and 3f) where continuous and dotted lines mark the same locations in both images. Additionally, we observe that the arcades exhibit consecutive maxima and minima (Figure 3g) distinguishing different edge positions of the visible half of the honeycomb. Interestingly, the lattice modulation does only depend on bias polarity, but not on the absolute value of the bias voltage, *i.e.* we observe very similar constant current images between + 0.1 V and + 1.5 V, respectively - 0.1 V and - 1.5 V, if we use the same tip. We checked about ten different locations on the $Bi_{14}Rh_3I_9$ surface using different micro-tips, always revealing the same lateral LDOS shift between occupied and unoccupied states. We also checked that thermal drift and piezo-creep are negligible by recording consecutive images at the same bias revealing lateral shifts below 0.5 Å between two consecutive images.

Figure 3h shows the honeycomb lattice (white dotted lines), the unit cell of the honeycomb lattice (gray rhomboid with unit vector $a_0$), and the unit cell of the apparent electronic modulation (black rectangle) superimposed on a constant-current image. This reveals a rectangular $\sqrt{3}\,a_0 \times a_0$ unit cell of the electronic structure with respect to the honeycomb unit cell of the 2DTI layer. This $\sqrt{3}\,a_0 \times a_0$ unit cell agrees with the a-b-plane of the trigonal unit cell of the 3D structure of $Bi_{14}Rh_3I_9$.[16] Since the appearance of the unit cell strongly depends on the sign of the bias (Figure 3a and 3d) and is partly tip-dependent (inset of Figure 2a), we exclude that it is given by atomic corrugations only. In contrast, we assume that it has mostly an electronic origin.

In order to elucidate this electronic effect and the bias-dependent lateral shift in more detail, we consider the specific orientation of the 2DTI honeycomb lattice on top of the underlying zigzag chains of the spacer layer. Figure 4a shows a sketch of the two layers on top of each other. It reveals three different stacking situations of the three-centered Bi atoms at the corners of the honeycomb lattice of the 2DTI layer (circles) with respect to the underlying spacer chains. In one situation, the triangular-prismatic voids line up perfectly with a spacer layer octahedron below (white circle), another where no direct neighbor from the spacer is present (black circle), and a third one where only a part of a spacer layer octahedron is found below (orange circle). The orange circles include two different stacking positions, both with a part of a spacer layer octahedron below, but both show very similar LDOS in simulated STM images (Figure 4c and 4d), such that they are not distinguished further. Below the main image of Figure 4a, two zooms are displayed which exhibit only one honeycomb. They are shifted with respect to each other by $\frac{\sqrt{3}}{2}a_0$ in horizontal direction and $a_0/2$ in vertical direction. This is exactly the shift of the arcade patterns between occupied and unoccupied states (Figure 3). It is likely, that the positions marked by the yellow circles are at the minima and the end points of the observed arcades (cf. height profile in Figure 3g), since these are the only three edge points of similar intensity in the STM images. This directly implies that they have the intermediate intensity within occupied and unoccupied states, which in both cases can act as a lever for the other two intensities. Thus, the other two positions marked by black and white circles must explain the lateral shift between occupied and unoccupied states. For the occupied states, one color (black or white) must identify the edge positions with lowest LDOS, while the other color identifies the edge positions with highest LDOS. The role of the colors must then be reversed for the unoccupied states.

Most probably, a hybridization of the Bi 6p orbitals of the 2DTI layer with the I 5p orbitals of the underlying octahedron is responsible for the contrast, since these orbitals exhibit the strongest DOS close to $E_F$ within their respective layers (Figure 4b). We believe that the LDOS for occupied states in vacuum decreases at positions where the trigonal-prismatic Bi atoms match with a spacer layer octahedron below (white circle). This suggestion is based on the reasonable assumption that a corresponding occupied state is a bonding state between Bi 6p and I 5p states located directly above each other, which leads to an increased LDOS between the layers and a decreased LDOS within the vacuum, if compared with a non-bonding state. In contrast, the unoccupied, antibonding state exhibits less LDOS between the layers and correspondingly more LDOS in vacuum. The same argument applies to the trigonal-prismatic Bi atoms on top of part of a spacer layer octahedron (orange circles), but with less strength, while the trigonal-prismatic Bi atoms above voids (black circle) should not exhibit a significant difference between the spatial distribution of occupied and unoccupied parts. This straightforwardly explains the observed contrast inversion, respectively the lateral shift of the arcades.

In order to further substantiate the above considerations, we use the DFT data from the layer model and extract spatially resolved density of states plots at a plane being 2.64 Å above the last Bi atoms of the surface 2DTI layer (Figure 4c and 4d). Notice that 2.64 Å is smaller than typical tip-surface distances implying better lateral resolution of the calculated data. Already at that distance, most of the intensity stems from the surface 2DTI layer with only a negligible signal from the underlying layers. To enhance the contrast, the square root of the real space LDOS is plotted.

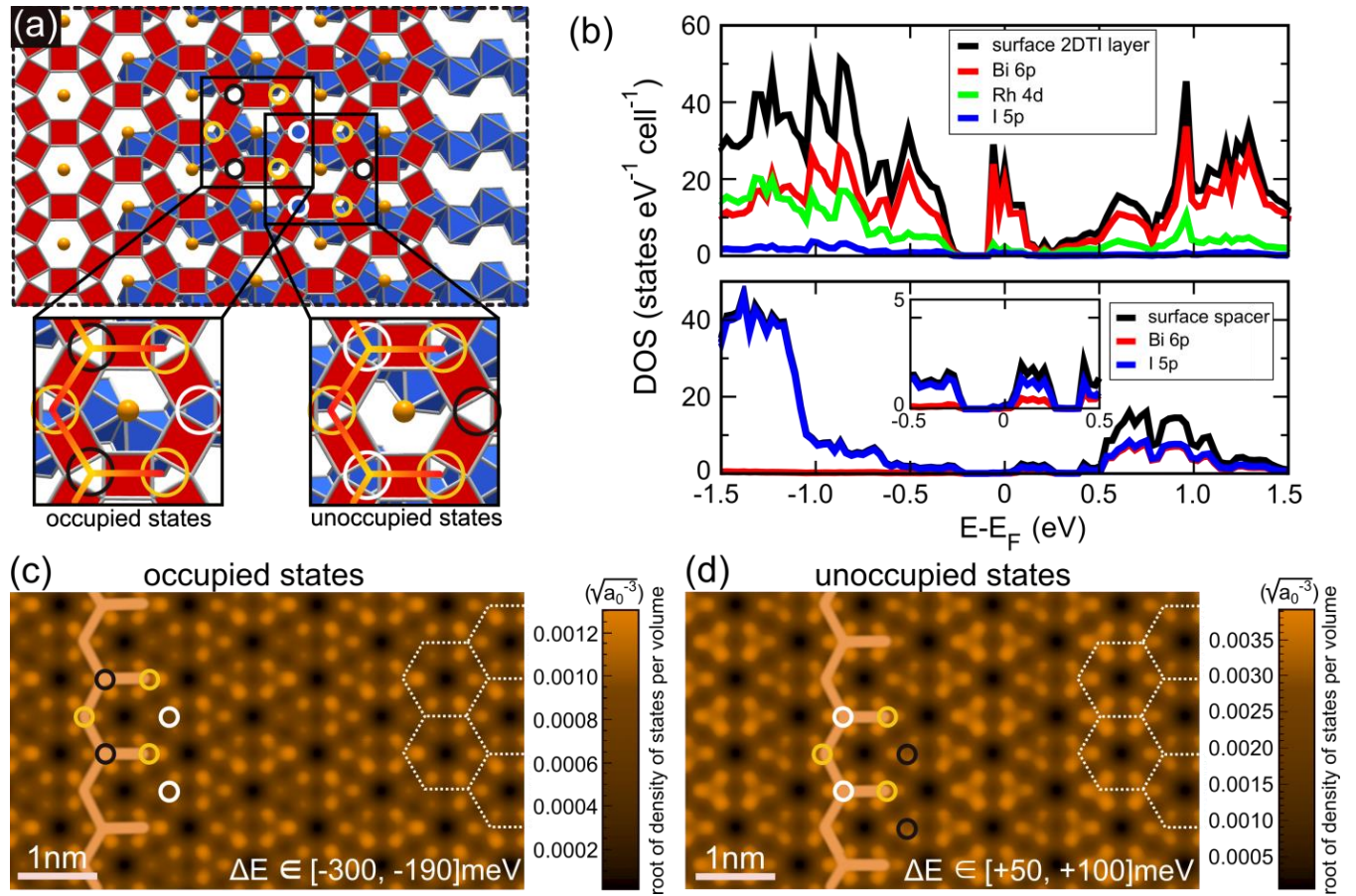

**Figure 4.** (a) Sketch of the surface 2DTI layer (red) with the underlying spacer layer (blue) including zooms in two honeycombs below as marked. The three different stacking situations of the three-centered Bi atoms with respect to the underlying spacer chains are marked by differently colored circles. White circles mark the positions where the triangular-prismatic voids line up perfectly with a spacer layer octahedron, dark circles mark the positions where no direct neighbor from the spacer is present, and orange circles indicate the situations where only a part of a spacer layer octahedron is present below the three-centered Bi atoms. The arcade structure observed in Figure 3 is superimposed in an orange-yellow color code within the zoomed images. Left (right) zoom marks the situation of the arcade for the occupied (unoccupied) states according to DFT as marked. (b) Calculated DOS (slab with iodine deficiency) projected to specific elemental orbitals of the surface 2DTI layer (top) and the surface spacer layer (bottom). Inset shows a close-up view for the region around $E_F$. The black lines mark the respective total layer DOS in both images. (c), (d) Calculated real space LDOS of $Bi_{14}Rh_3I_9$ for the same spatial area as in (a) at a plane being 2.64 Å above the surface 2DTI layer and integrated over (c) an energy range within the occupied states ($E \in$ [-300, -190] meV) and (d) an energy range within the unoccupied states ($E \in$ [+50, +100] meV). The data are deduced from the calculation of the finite four layer slab without iodine deficiency. The bright dots appear at the positions of the Bi atoms which form the triangular-prismatic voids of the 2DTI layer. A regular honeycomb lattice is superimposed as a white dotted line pattern. Contrast modulations at the triangles are visible with superimposed continuous line patterns (guide to the eye) matching the brighter ones. Note the shift of the line pattern between (c) and (d). Colored circles are drawn at identical positions as in (a).

The resulting patterns for occupied and unoccupied states are shown in Figure 4c and 4d, respectively. The bright dots in the calculated images are located above the Bi atoms which form the triangular-prismatic voids at the corners of the honeycomb lattice. The DOS between the Bi atoms around the voids is more intense than the DOS at the Bi-Rh bonds within the Rh centered cubes, probably due to the larger distance of the Rh atoms from the surface and the dominating contribution of the Bi 6p states to the total DOS of the 2DTI layer with respect to the Rh and I states (top panel in Figure 4b). Interestingly, one triangle of Bi-atoms within the honeycomb appears significantly weaker than the others as marked by a white (black) circle in Figure 4c(d). These specific triangular-prismatic voids are at the positions marked by the same colored circles in Figure 4a. The same arcade pattern as in the insets of Figure 4a is also overlaid. In line with the experimental findings, the position of the weak triangles is shifted between occupied (Figure 4c) and unoccupied (Figure 4d) states. We checked that this energy dependent shift is visible throughout the whole vacuum section up to 4.1 Å, where the signal gets too weak with respect to the numerical noise. Thus, even though the corrugation within the arcade pattern (inset Figure 3f) is not reproduced, the DFT data qualitatively agree with the experimental findings by reproducing the points of strongly reduced DOS. Quantitatively, the lattice modulation measured in the experiment is significantly stronger than the modulation found in the DFT calculation, most probably due to relaxation effects between the layers, which are not included in the calculation. Nevertheless, we conclude that the $\sqrt{3}\,a_0 \times a_0$ electronic structure of the surface 2DTI layer is naturally induced by the underlying spacer layer.

We also find a bias-dependent atomic appearance of the surface spacer layer. The atomically resolved STM images in Figure 3b and 3e display the iodine atoms of the upper part of the spacer layer (cf. model in Figure 1a and 1b), since these atoms are closest to the tip and, moreover, exhibit by far the highest LDOS within the spacer layer (Figure 4b). We find a rather regular hexagonal pattern except of the missing iodine atoms within the occupied states, but a more triangularly shaped appearance within the unoccupied states. The explanation can be deduced from the LDOS projected to the individual orbitals (bottom panel in Figure 4b). At the occupied states (-1.3 eV to $E_F$), the I 5p states largely exceed the contribution from the Bi 6p states. In contrast, at the unoccupied states ($E_F$ to +0.5 eV), finite contributions of both the I 5p and the Bi 6p states are discernible. Thus, the lack of Bi 6p contributions within the occupied states most probably prevents the visibility of underlying Bi-atoms and therefore only the hexagonal arrangement of the iodine ions is visible. Oppositely, at the unoccupied states, the triangular appearance evolves from the underlying Bi-atoms, as iodine ions that belong to the same $BiI_6$ octahedra (depicted as white triangles in Figure 1b) become grouped.

CONCLUSIONS:

In summary, our combined experimental and DFT analysis of the weak topological insulator $Bi_{14}Rh_3I_9$ reveals that the observed band shift of about 100 - 200 meV is due to the unavoidable presence of polar surfaces after cleavage. This band shift is only found at the topologically non-trivial cationic surface, where the surface-induced dipole is compensated by electronic reconstruction. In contrast, surface reconstruction by iodine evaporation takes place at the surface spacer layer which shifts the gap back to its bulk position at the Fermi level. Since the potential shift of the terminating 2DTI layer is already well screened at the first inner 2DTI layer, an experimental strategy to move the topologically protected edge state to $E_F$ could simply leave the spacer layer on top and expose the edge of the underlying 2DTI layer by adequate surface scratching.

In addition, we found that the coupling of the 2DTI layer to the underlying spacer layer destroys the honeycomb symmetry of the 2DTI electronic structure, leading to a $\sqrt{3}\, a_0 \times a_0$ unit cell of the surface 2DTI layer with respect to the honeycomb structure in STM images. Consequently, STM reveals fingerprints of the structural arrangement of the underneath spacer layer.

METHODS

**Experimental methods:** The scanning tunneling microscopy measurements have been performed in a home-built scanning tunneling microscope in UHV ($p_{base}$ = $10^{-11}$ mbar) at a temperature of $T$ = 6 K.[29] Topographic STM images and *dI/dV* images are recorded in constant-current mode with voltage *V* applied to the sample. The *dI/dV(V)* spectra are measured after stabilizing the tip at a sample voltage $V_{stab}$ and a current $I_{stab}$ before opening the feedback loop. The spectroscopic measurements are carried out using a lock-in technique with a modulation frequency $v$ = 1.4 kHz and amplitude $V_{mod}$ = 4 mV resulting in an energy resolution $\delta E \approx 7$ meV.[30] To first order, $dI/dV(V)$ is proportional to the local density of states (LDOS) of the actual sample position at energy $E$ with respect to the Fermi level $E_{F,sample}$ of the sample ($E = E_{F,sample} + eV$) multiplied by the LDOS of the tip at the Fermi level $E_{F,tip}$ of the tip.

The angle-resolved photoemission spectroscopy measurements have been performed at the $1^3$-ARPES facility [31] using the synchrotron radiation from the BESSY storage ring. The sample temperature has been kept below 1 K during these measurements.

The single crystal $Bi_{14}Rh_3I_9$ samples have been prepared in Dresden according to a previously described method.[16,18] Prior to the STM and ARPES measurements, the samples were cleaved *in-situ* at room temperature using a commercial copper tape, leading to atomically flat terraces.

**Computational details:** All density functional theory calculations were done with the Full-Potential Local-Orbital (FPLO) code [32] using the local density approximation (LDA) in the PW92-parameterization.[33] The linear tetrahedron method with a 12 × 7 × 7 $k$-mesh was employed for the self-consistent bulk $Bi_{14}Rh_3I_9$ calculations, and with 6 × 4 × 1 $k$-points for the four-layer stack. The self-consistent calculations were carried out with SO coupling included. This effort is needed, since the gap is only opened by SOC. The following basis states are treated as valence states: Bi: 5s, 5p, 5d, 6s, 7s, 6p, 7p, 6d; Rh: 4s, 4p, 5s, 6s, 4d, 5d, 5p; I: 4s, 4p, 4d, 5s, 6s, 5p, 6p, 5d. The considered layer stack has the same lateral cell dimensions, atomic positions, and space group as the bulk, but two elementary cells in stack direction, separated from the next stack by a vacuum layer of about 28 Å thickness. Thus, the elementary cell of the slab contains four chemical units (104 atoms).

AUTHOR INFORMATION

**Corresponding Author**

mmorgens@physik.rwth-aachen.de

ACKNOWLEDGMENT

We acknowledge the help of Volodymyr Zabolotnyy, Danil Evtushinsky and Setti Thirupathaiah at the beamline and the grants by the German science foundation *via* Mo 858/13-1, IS 250/1-1 and RI 932/7-1 being part of the Priority Programme "Topological Insulators" (SPP 1666). C. P. thanks the Fonds National de la Recherche Luxembourg for funding.